\documentclass[%
 aip,jcp,
 amsmath,amssymb,
 preprint,%
]{revtex4-1}

\usepackage{graphicx}
\usepackage{dcolumn}
\usepackage{bm}

\usepackage[utf8]{inputenc}
\usepackage[T1]{fontenc}
\usepackage{mathptmx}
\usepackage{etoolbox}
\usepackage{threeparttable}
\usepackage[section]{placeins}
\usepackage{longtable}
\usepackage{multirow}
\usepackage{booktabs}

\makeatletter
\def\@email#1#2{%
 \endgroup
 \patchcmd{\titleblock@produce}
  {\frontmatter@RRAPformat}
  {\frontmatter@RRAPformat{\produce@RRAP{*#1\href{mailto:#2}{#2}}}\frontmatter@RRAPformat}
  {}{}
}%
\makeatother

\begin{document}

\preprint{AIP/123-QED}

\title{Combining Localized Orbital Scaling Correction and Bethe-Salpeter Equation for Accurate Excitation Energies}

\author{Jiachen Li}
\affiliation{Department of Chemistry, Duke University, Durham, NC 27708, USA}
\author{Ye Jin}
\affiliation{Department of Chemistry, Duke University, Durham, NC 27708, USA}
\author{Neil Qiang Su}
\affiliation{Department of Chemistry, Duke University, Durham, NC 27708, USA}
\author{Weitao Yang}
\affiliation{Department of Chemistry, Duke University, Durham, NC 27708, USA}
\email{weitao.yang@duke.edu}

\date{\today}

\begin{abstract}
We applied localized orbital scaling correction (LOSC) in Bethe-Salpeter equation (BSE) to predict accurate excitation energies for molecules.
LOSC systematically eliminates the delocalization error in the density functional approximation and is capable of approximating quasiparticle (QP) energies with accuracy similar or better than the $GW$  Green's function approach and with much less computational cost.
The QP energies from LOSC instead of commonly used $G_{0}W_{0}$ and ev$GW$ are directly used in BSE.
We show that the BSE/LOSC approach greatly outperforms the commonly used BSE/$G_{0}W_{0}$ approach for predicting excitations with different characters.
For the calculations for Truhlar-Gagliardi test set containing valence,
charge transfer (CT) and Rydberg excitations,
BSE/LOSC with the Tamm-Dancoff approximation provides a comparable accuracy to time-dependent density functional theory (TDDFT) and BSE/ev$GW$.
For the calculations of Stein CT test set and Rydberg excitations of atoms,
BSE/LOSC considerably outperforms both BSE/$G_{0}W_{0}$ and TDDFT approaches with a reduced starting point dependence.
BSE/LOSC is thus a promising and efficient approach to calculate excitation energies for molecular systems.
\end{abstract}

\maketitle

\section{\label{sec:introduction}INTRODUCTION}
Electronic excitation energy is one of the most important quantities for describing the electronic excited states.
It can be compared with the 0-0 energy, which is measured by the optical spectroscopy.
Computationally, determining the excitation energy from the electronic structure theory plays a critical role for obtaining
insights about various phenomena and processes in chemistry, biochemistry and material science,
such as molecular drug delivery\cite{velemaOpticalControlAntibacterial2013,beharryAzobenzenePhotoswitchingUltraviolet2011}
and solar cells\cite{greggExcitonicSolarCells2003,peumansEfficientBulkHeterojunction2010,hagfeldtMolecularPhotovoltaics2000}.
In past decades,
many efforts have been devoted to develop accurate and efficient theoretical
approaches to compute excitation energies.
One of the most popular approaches is time-dependent density functional theory (TDDFT)\cite{rungeDensityFunctionalTheoryTimeDependent1984,casidaTimeDependentDensityFunctional1995,ullrichTimeDependentDensityFunctionalTheory2011}.
TDDFT has been widely implemented in modern quantum chemistry packages to calculate
energies, structures and other properties of excited states for molecular and periodic systems\cite{casidaTimedependentDensityfunctionalTheory2009,casidaLinearResponseTimeDependentDensity2006,yuanLinearResponseApproach2009,laurentTDDFTBenchmarksReview2013}.
The success of TDDFT stems from the good compromise between the accuracy and the computational cost.
However, TDDFT still suffers from several problems.
For example,
it is known that TDDFT with commonly used density functional approximations (DFAs)
fails to describe Rydberg excitations and charge transfer (CT) excitations\cite{laurentTDDFTBenchmarksReview2013,kaurWhatAccuracyLimit2019}.
This issue can be attributed to the incorrect description for the long-range
behavior for the potential energy surface\cite{dreuwLongrangeChargetransferExcited2003,tozerRelationshipLongrangeChargetransfer2003}.
Efforts for correcting the long-range behavior includes using range-separated or
Coulomb-attenuated functionals\cite{leiningerCombiningLongrangeConfiguration1997,besleyTimedependentDensityFunctional2009,peachExcitationEnergiesDensity2008}
and mixing the Hartree-Fock (HF)\cite{slaterNoteHartreeMethod1930,szaboModernQuantumChemistry2012} exchange in DFAs\cite{zhaoDensityFunctionalSpectroscopy2006}.
Besides the failures for describing CT and Rydberg excitations,
TDDFT has an undesired dependence on the exchange-correlation (XC) kernel stemming from different choices of DFAs.
The difference originating from using different DFAs can exceed $1.0$
\,{eV} for valence excitation energies and even exceed $2.0$ \,{eV} for Rydberg
excitation energies\cite{kaurWhatAccuracyLimit2019,laurentTDDFTBenchmarksReview2013}. \\

Recently, Bethe-Salpeter equation (BSE)\cite{shamManyParticleDerivationEffectiveMass1966,salpeterRelativisticEquationBoundState1951,hankeManyParticleEffectsOptical1979,blaseBetheSalpeterEquation2020,martinInteractingElectrons2016,onidaElectronicExcitationsDensityfunctional2002}
in the Green's function many-body perturbation theory\cite{hedinNewMethodCalculating1965,martinInteractingElectrons2016}
has gained increasing attention to compute the optical spectroscopy for
molecular systems.
The BSE approach commonly takes the energy levels computed from the $GW$
approximation as the input\cite{hedinNewMethodCalculating1965,martinInteractingElectrons2016,golzeGWCompendiumPractical2019,reiningGWApproximationContent2018}
and this approach is denoted as the BSE/$GW$ approach.
In the BSE/$GW$ approach,
the screened interaction is used instead of the bare Coulomb interaction to
describe the electron-hole interaction.
The screened interaction is formulated with the quasiparticle (QP) energies from the $GW$ calculation.
It is known that the QP energies from $GW$ are more accurate for predicting
HOMO-LUMO gap (fundamental gap) than the conventional Kohn-Sham (KS) orbital energies.
Besides the improved gap prediction,
energies also have more clear physical meanings that they are
interpreted as the charged excitation energies,
or excitation energies for electron removal and addition.
It has been shown that the $GW$ approximation substantially improves the
accuracy of predicting energy levels over the KS density functional theory
(KS-DFT) approach for both occupied and unoccupied states\cite{vansettenGW100BenchmarkingG0W02015,martinInteractingElectrons2016,kaplanQuasiParticleSelfConsistentGW2016},
which are the key quantities to calculate the excitation energies.
Because the correct long-range behavior in the BSE/$GW$ approach and the
importance of the dynamical screening in real systems,
BSE/$GW$ has been applied to calculate excitation energies for systems of
different sizes\cite{azariasBetheSalpeterStudyCationic2017,azariasCalculationsTransitionEnergies2017,escuderoModelingPhotochromeTiO22017,jacqueminBenchmarkBetheSalpeterTriplet2017,jacqueminBetheSalpeterFormalism2017,blaseChargetransferExcitationsMolecular2011,ziaeiGWBSEApproachS12016,jacqueminBenchmarkingBetheSalpeter2015,faberExcitedStatesProperties2014,jiangRealtimeGWBSEInvestigations2021,liuAllelectronInitioBetheSalpeter2020,rinkeFirstPrinciplesOpticalSpectra2012,dvorakQuantumEmbeddingTheory2019,albrechtInitioCalculationExcitonic1998,cudazzoExcitonBandStructure2016,romanielloDoubleExcitationsFinite2009,disabatinoScrutinizingGWBasedMethods2021}.
However, the BSE/$GW$ approach still has several challenges.
First, although BSE has the same scaling as TDDFT\cite{krauseImplementationBetheSalpeter2017,ghoshConceptsMethodsModern2016},
which is $\mathcal{O}(N^4)$ ($N$ is the size of the system),
the preceding $GW$ calculation is computationally expensive.
In the fully analytical treatment of $GW$,
the scaling of solving the random phase approximation (RPA) equation is
$\mathcal{O}(N^6)$ and the scaling of evaluating the self-energy is $\mathcal{O}(N^5)$\cite{golzeGWCompendiumPractical2019,vansettenGWMethodQuantumChemistry2013}.
Thus, the computational-demanding $GW$ calculation is the bottleneck of the
BSE/$GW$ approach.
To reduce the cost of $GW$ calculations,
different techniques can be used.
For example,
the cost of formulating the response function can be reduced to $\mathcal{O}(N^4)$ by using the plasmon-pole models\cite{deslippeBerkeleyGWMassivelyParallel2012} or the analytic continuation\cite{vansettenGW100BenchmarkingG0W02015,golzeGWCompendiumPractical2019,shirleyCorePolarizationSemiconductors1992}.
The cost of evaluating the self-energy can be reduced to $\mathcal{O}(N^5)$ in the contour deformation approach\cite{golzeCoreLevelBindingEnergies2018} and $\mathcal{O}(N^4)$ in the analytical continuation approach\cite{ducheminRobustAnalyticContinuationApproach2020}.
Recently the cubic scaling implementations\cite{wilhelmGWCalculationsThousands2018,ducheminCubicScalingAllElectronGW2021} of $GW$ calculations have also gained increasing attention.
Second, the performance the most used BSE/$G_0W_0$ approach strongly depends on
the choice of the DFA.
Because of the perturbative nature of the one-shot $G_0W_0$ method,
the accuracy of $G_0W_0$ strongly depends on the starting point\cite{keAllelectronGWMethods2011,maromBenchmarkGWMethods2012,fuchsQuasiparticleBandStructure2007}.
This undesired dependence is inherited in the BSE/$G_0W_0$ approach.
It has been shown that the accuracy of BSE/$G_0W_0$ for predicting excitation
energies of molecular systems is largely affected by the starting point\cite{jacqueminBenchmarkingBetheSalpeter2015}.
$G_0W_0$ based on range-separated functionals and tuned hybrid functionals provides more accurate QP energies\cite{dauthPiecewiseLinearityGW2016,korzdorferAssessmentPerformanceTuned2012,hollasModelingLiquidPhotoemission2016,golzeCoreLevelBindingEnergies2018,brunevalSystematicBenchmarkInitio2015},
which lead to better excitation energies in BSE/$G_0W_0$.
It has been shown that BSE/$G_0W_0$ with optimally tuned hybrid functionals and range-separated functionals predicts accurate core electron excitation energies and low-lying excitation energies\cite{moninoSpinConservedSpinFlipOptical2021,yaoAllElectronBSEGW2022}.
One path to improve the accuracy is to introduce the self-consistency into the $GW$ calculations.
BSE combining with the eigenvalue self-consistent $GW$ (ev$GW$) approach\cite{kaplanQuasiParticleSelfConsistentGW2016}
is shown to predict accurate excitation energies for organic molecules and CT
systems with the reduced starting point dependence\cite{jacqueminBenchmarkingBetheSalpeter2015,jacqueminAssessmentConvergencePartially2016,jacqueminBenchmarkBetheSalpeterTriplet2017,blaseChargetransferExcitationsMolecular2011}.
It has been shown that BSE/ev$GW$ provides comparable accuracy to TDDFT with hybrid functionals and the difference of excitation energies originating from using different DFAs in BSE/ev$GW$ is only around $0.1$ to $0.3$ \,{eV}\cite{jacqueminBenchmarkBetheSalpeterTriplet2017,jacqueminAssessmentConvergencePartially2016,jacqueminBetheSalpeterFormalism2017}.
In practice,
few iterations are necessary for ev$GW$ calculations to reach the convergence\cite{kaplanQuasiParticleSelfConsistentGW2016}.
The additional computational cost in ev$GW$ is only prohibited for large systems. \\

While Green's function theory provides QP energies by construction,
there are parallel development in DFT.
Within DFT,
the physical meaning of the one-electron orbital energies of the frontier orbitals,
namely HOMO and LUMO,
has been established based on the three key theoretical results.
First,
the Janak theorem links Kohn-Sham orbital energies to the derivatives of the total energy with respect to the orbital occupation numbers,
which are, however,
not related to any physical observables directly\cite{janakProofThatFrac1978}.
Second, the derivatives of the total energy with respect to the total electron number,
the chemical potentials,
are respectively the negative of the first ionization potential (IP) and the first electron affinity (EA) for the exact functional based on the linear condition on the behavior of energy for fractional charges\cite{perdewDensityFunctionalTheoryFractional1982,perdewExchangeCorrelationOpen2007,yangDegenerateGroundStates2000}.
Third, the chemical potentials were established to be equal to the derivatives of the total energy with respect to the HOMO/LUMO orbital occupation numbers in the Kohn-Sham calculation with XC energy being functionals of the density,
or the generalized Kohn-Sham calculation with XC energy being functionals of the noninteracting one-electron density matrix\cite{cohenFractionalChargePerspective2008}.
Combining these theoretical results,
the HOMO and LUMO energies are thus the approximation to the negative of the first IP and the first EA,
respectively, as first established by Cohen et. al\cite{cohenFractionalChargePerspective2008}.
This interpretation of the frontier orbital energies has been further extended for other orbitals:
Kohn-Sham or generalized Kohn-Sham orbital energies are corresponding QP energies,
from the DFA used\cite{meiApproximatingQuasiparticleExcitation2019}.
This extension was based on extensive observation for a large set of molecules that DFAs,
which were designed with minimal delocalization error and provide accurate prediction of IP and EA from the HOMO and LUMO energies,
also predict other QP energies from the corresponding generalized Kohn-Sham orbital energies with similar excellent accuracy as HOMO and LUMO orbitals.
Therefore, accurate QP energies can be provided from the ground state calculations of DFT\cite{meiApproximatingQuasiparticleExcitation2019}. \\

We want to leverage these recent developments of DFT within the BSE formalism for electronic excitations to achieve and improve the accuracy of the BSE/$G_{0}W_{0}$ approach at more affordable computational cost, based on the localized orbital scaling correction (LOSC)\cite{liLocalizedOrbitalScaling2018}.
Let us briefly review the development of LOSC and other DFAs.
Over past decades,
DFT\cite{hohenbergInhomogeneousElectronGas1964,kohnSelfConsistentEquationsIncluding1965,parrDensityFunctionalTheoryAtoms1989}
has become the most popular tool in the electronic structure theory.
In DFT,
the complicated electron correlation effect can be properly and efficiently
described by DFAs,
including local density approximations (LDAs)\cite{barthLocalExchangecorrelationPotential1972,voskoAccurateSpindependentElectron1980},
generalized gradient approximations (GGAs)\cite{beckeDensityfunctionalExchangeenergyApproximation1988,leeDevelopmentColleSalvettiCorrelationenergy1988,perdewAccurateSimpleAnalytic1992}
and hybrid functionals\cite{beckeDensityfunctionalExchangeenergyApproximation1988,beckeNewMixingHartree1993}.
However, the predictive power of DFT is impaired by some intrinsic deficiencies.
It has been shown that the delocalization error\cite{cohenInsightsCurrentLimitations2008,mori-sanchezLocalizationDelocalizationErrors2008}
is responsible for many failures in DFT,
such as the band-gap prediction\cite{mori-sanchezLocalizationDelocalizationErrors2008}.
The delocalization error in mainstream DFAs is manifested in small molecules as the violation of the
Perdew–Parr–Levy–Balduz (PPLB) condition\cite{perdewDensityFunctionalTheoryFractional1982,perdewExchangeCorrelationOpen2007,yangDegenerateGroundStates2000}
showing that the total energy of a system as a function of the electron number
should be piecewise linear between energies at integer points.
In 2011, global scaling correction (GSC)\cite{zhengImprovingBandGap2011} was
developed to impose the PPLB condition by using canonical occupations and
curvatures that are constructed from canonical orbitals.
It has shown that GSC largely restores the linearity behavior and predicts
accurate band gaps for systems of all sizes.
However, GSC offers zero corrections to total energies at integers,
which implies the correction is not size-consistent.
To provide a size-consistent correction,
local scaling correction (LSC)\cite{zhengScalingCorrectionApproaches2015,liLocalScalingCorrection2015}
was developed by using local fractional information.
But LSC has numerical difficulties for capturing tiny fractions\cite{zhengScalingCorrectionApproaches2015,liLocalizedOrbitalScaling2018}.
To combine merits of describing global fractions in GSC and local fractions in LSC,
LOSC was developed to systematically eliminate the delocalization error in a
size-consistent manner by utilizing orbitalets\cite{liLocalizedOrbitalScaling2018}.
Orbitalets are defined as a set of orbitals localized in both physical and
energy spaces that are obtained by the restrained Boys localization\cite{liLocalizedOrbitalScaling2018}.
By using orbitalets,
the LOSC correction can be applied to the global and local regions of the system
in a dynamical way.
It has been shown that LOSC successfully describes dissociation of cationic
species, band gaps and photoemission spectrum\cite{liLocalizedOrbitalScaling2018,meiApproximatingQuasiparticleExcitation2019}.
As shown in the recent work from our group,
accurate QP energies can be approximated from LOSC\cite{meiApproximatingQuasiparticleExcitation2019}.
LOSC provides very similar or better photoemission spectrums and fundamental gaps to those
obtained from the fully self-consistent $GW$ (sc$GW$) approach\cite{meiApproximatingQuasiparticleExcitation2019}.
As shown in Section.5 in the Supporting Information,
$G_0W_0$ with conventional DFAs gives underestimated fundamental gaps compared with ev$GW$.
LOSC provides larger fundamental gaps,
which are similar to the ev$GW$ level.
In addition,
Fundamental gaps obtained from LOSC with hybrid functionals are closer to those obtained from ev$GW$ than LOSC with GGA functionals.
Furthermore, there are further development of LOSC approach\cite{suPreservingSymmetryDegeneracy2020}
and a open-source software\cite{meiLibSCLibraryScaling2022}.
Because LOSC is computationally favorable\cite{liLocalizedOrbitalScaling2018},
it is a promising alternative to the $GW$ methods in BSE/$GW$.
In this work we introduced the BSE/LOSC approach,
which directly uses LOSC orbital energies in BSE to calculate excitation energies.
Applying LOSC in BSE to bypass the $GW$ calculation shares a similar thinking as combining Koopmans-compliant functionals with BSE in Ref.\citenum{elliottKoopmansMeetsBethe2019}.
In the KI-BSE approach,
the QP energies are derived from Koopmans-compliant functionals and the screened interaction is obtained via a direct minimization on top of a maximally localized Wannier function basis\cite{elliottKoopmansMeetsBethe2019}.
It is shown that KI-BSE provides similar accuracy to BSE/$G_0W_0$\cite{elliottKoopmansMeetsBethe2019}.
We show that the BSE/LOSC approach considerably outperforms BSE/$G_0W_0$ for
predicting valence, Rydberg and CT excitation energies with lower computational
cost. \\

\section{\label{sec:theory}THEORY}

\subsection{\label{subsec:losc}Localized orbital scaling correction}
The LOSC correction to the total energy is expressed as\cite{liLocalizedOrbitalScaling2018}
\begin{equation}\label{eq:losc_energy}
    \Delta E^{\text{LOSC}} =
    \sum_{pq} \frac{1}{2} \kappa_{pq}\lambda_{pq}
    (\delta_{pq} - \lambda_{pq}) \text{,}
\end{equation}
where $\lambda$ is the local occupation matrix and $\kappa$ is the curvature matrix.
In Eq.\ref{eq:losc_energy} and following equations,
we use $i$, $j$ for occupied orbitals,
$a$, $b$ for virtual orbitals and $p$, $q$ for general orbitals. \\

The local occupation matrix $\lambda$ in Eq.\ref{eq:losc_energy} is defined as\cite{liLocalizedOrbitalScaling2018}
\begin{equation}\label{eq:lamda_matrix}
  \lambda_{pq}= \langle \phi_p |\rho_s| \phi_q \rangle \text{,}
\end{equation}
where $\rho_s$ is the KS density matrix and $\{ \phi_p \}$ is the set of orbitalets.
The diagonal elements of $\lambda$ contain the information of the fractional
electron distribution and the off-diagonal elements bring corrections to the
unphysical interaction between the local fractions centered at different
positions\cite{liLocalizedOrbitalScaling2018}. \\

The curvature matrix $\kappa$ in Eq.\ref{eq:losc_energy} is defined as\cite{liLocalizedOrbitalScaling2018}
\begin{equation}\label{eq:kappa_matrix}
    \begin{split}
        \frac{1}{2} \kappa_{pq} =&
        \frac{1}{2} \iint \frac{\rho_p(r) \rho_q(r')}{|r-r'|} dr dr'
        - \frac{ \tau C_x }{3} \int [\rho_p(r)]^{\frac{2}{3}}
        [\rho_q(r)]^{\frac{2}{3}} dr \text{,}
    \end{split}
\end{equation}
where $\rho_p(r)=|\phi_q(r)|^2$ is the density of the orbitalet,
$C_x=\frac{3}{4}(\frac{6}{\pi})^{1/3}$ and
$\tau = 1.2378$\cite{liLocalizedOrbitalScaling2018,zhengImprovingBandGap2011}.
Eq.\ref{eq:kappa_matrix} exactly compensates the deviation from the linearity
condition under the frozen-orbital assumption\cite{liLocalizedOrbitalScaling2018}.\\

Orbitalets $\{ \phi_p \}$ used in $\lambda$ and $\kappa$ are obtained the
restrained Boys localization,
which minimizes the following spread function\cite{liLocalizedOrbitalScaling2018}
\begin{equation}\label{eq:boys}
    F = \sum_p \bigg[\langle \phi_p |r^2|\phi_p \rangle -
    \langle \phi_p |r|\phi_p \rangle^2 \bigg] +
    \sum_{pq}w_{pq}|U_{pq}|^2 \text{,}
\end{equation}
where $w$ is the penalty function defined in Ref.\citenum{liLocalizedOrbitalScaling2018}
and $U$ is the unitary transformation matrix from canonical orbitals to orbitalets.
This restrained Boys localization mixes both occupied and virtual orbitals,
which is different from the original Boys localization that only mixes occupied
orbitals\cite{fosterCanonicalConfigurationalInteraction1960}.
In Eq.\ref{eq:boys},
the first term that has the similar form as the spread function in the original
Boys localization ensures orbitals localized in the physical space,
and the second term forbids mixing between canonical orbitals that are far apart
in the energy space by using the penalty function $w$.
More recently, a modified localization function was developed which leads to similar accuracy,
but preserving the degeneracy of orbitals\cite{suPreservingSymmetryDegeneracy2020}.
We have used the localization function of Eq.\ref{eq:boys} in this work.\\

With the LOSC correction defined in Eq.\ref{eq:losc_energy} the total energy is
expressed as
\begin{equation}
    E= E^{\text{DFA}} + \Delta E^{\text{LOSC}} \text{.}
\end{equation}
And the LOSC correction to the Hamiltonian is
\begin{equation} \label{eq:losc_h}
    H = H^{\text{DFA}} + \Delta h^{\text{LOSC}} \text{,}
\end{equation}
where
\begin{equation}
    \Delta h^{\text{LOSC}} = \sum_p \kappa_{pp} (\frac{1}{2} - \lambda_{pp})
    |\phi_p \rangle \langle \phi_p | -
    \sum_{p \neq q} \kappa_{pq} \lambda_{pq} |\phi_p \rangle \langle \phi_q | \text{.}
\end{equation}
The LOSC orbital energies are obtained by diagonalizing $H$ in Eq.\ref{eq:losc_h}. \\

The LOSC correction can be applied in either the
self-consistent manner (SCF LOSC) or the post-SCF manner (post-SCF LOSC).
As shown in Ref.\citenum{liLocalizedOrbitalScaling2018},
SCF LOSC provides improved orbitals and corrects the wrong electron density caused by the delocalization error.
As shown in Table S18 in Section.6 in the Supporting Information,
dipole moments of tested molecules obtained from SCF LOSC are similar to those obtained from quasiparticle-self-consistent $GW$ (qs$GW$) and sc$GW$.
This indicates that the electron density from LOSC is similar to those from qs$GW$ and sc$GW$,
which are also more localized compared with the electron density from KS-DFT\cite{kaplanQuasiParticleSelfConsistentGW2016,carusoFirstprinciplesDescriptionCharge2014}.
SCF LOSC with hybrid functionals provides better agreement with $GW$ results,
which is similar to fundamental gaps.
However,
as discussed in Ref.\citenum{kaplanQuasiParticleSelfConsistentGW2016} and shown in numerical results in Table.S2 and Table.S7 in the Supporting Information,
orbital updates have a minor effect on QP energies for the systems studied in this work.
Therefore, we do not expect SCF to change the results.
In addition to the minor effects from using SCF LOSC orbitals,
as shown in numerical results of Table.S2 and Table.S7 in the Supporting Information,
SCF LOSC has convergence problems in previous implementation\cite{liLocalizedOrbitalScaling2018} when using the augmented basis sets or
calculating large systems,
although the convergence problem can be overcome with the recently developed SCF method for LOSC\cite{meiApproximatingQuasiparticleExcitation2019}.
Thus we focus on the simplest BSE/post-SCF LOSC approach,
denoted as BSE/LOSC. \\

\subsection{\label{subsec:bse}The Bethe-Salpeter equation}
The key idea of this work is to use LOSC orbital energies
$\{\varepsilon^{\text{LOSC}}_p\}$ as the input in BSE.
With the static approximation for the screened interaction\cite{krauseImplementationBetheSalpeter2017,blaseBetheSalpeterEquation2020,ghoshConceptsMethodsModern2016},
the working equation of BSE is a generalized eigenvalue equation\cite{krauseImplementationBetheSalpeter2017,ghoshConceptsMethodsModern2016,blaseBetheSalpeterEquation2020},
which is similar to the Casida equation in TDDFT\cite{ullrichTimeDependentDensityFunctionalTheory2011,casidaTimeDependentDensityFunctional1995}
\begin{equation}\label{eq:bse}
    \begin{bmatrix}
        \mathbf{A} & \mathbf{B} \\
        \mathbf{B^*} & \mathbf{A^*}
    \end{bmatrix}
    \begin{bmatrix}
        \mathbf{X} \\
        \mathbf{Y}
    \end{bmatrix}
    = \omega
    \begin{bmatrix}
        \mathbf{I} & \mathbf{0} \\
        \mathbf{0} & \mathbf{-I}
    \end{bmatrix}
    \begin{bmatrix}
        \mathbf{X} \\
        \mathbf{Y}
    \end{bmatrix} \text{,}
\end{equation}
where $\omega$ is the excitation energies.
The $\mathbf{A}$, $\mathbf{B}$ matrices in Eq.\ref{eq:bse} are defined as
\begin{align}
    A_{ia,jb} &= \delta_{ij} \delta_{ab} (\varepsilon_a
    - \varepsilon_i) + v_{ia,jb} - W_{ij,ab} \label{eq:a} \text{,}\\
    B_{ia,jb} &= v_{ia,bj} - W_{ib,aj} \label{eq:b} \text{,}
\end{align}
where $\{ \varepsilon_p \}$ are input orbital energies.
In Eq.\ref{eq:bse}, $v$ is the Coulomb interaction defined as
\begin{equation}
    v_{pq,rs} =
    \int dx_{1}dx_{2}\frac{\varphi_{p}^{*}(x_{1})\varphi_{r}^{*}(x_{2})\varphi_{q}(x_{1})\varphi_{s}(x_{2})}{|\mathbf{r_{1}}-\mathbf{r_{2}}|} \text{,}
\end{equation}
where $\{ \varphi_p \}$ is the set of input orbitals.
And $W$ is the screened interaction defined as
\begin{equation}
    W_{pq,rs} = \sum_{tu} (\epsilon^{-1})_{pq,tu} v_{tu,rs} \text{,}
\end{equation}
where the dielectric function is calculated by the static response function $\chi$\cite{krauseImplementationBetheSalpeter2017,ghoshConceptsMethodsModern2016}
\begin{align}
    \epsilon_{pq,rs} &= \delta_{pr}\delta_{qs} - v_{pq,rs}\chi_{rs,rs} \text{.} \\
    \chi_{ia,ia} &= \chi_{ai,ai} =(\varepsilon_i - \varepsilon_a)^{-1}
\end{align}

The Tamm-Dancoff approximation (TDA) in BSE is obtained by neglecting the $\mathbf{B}$
matrix in Eq.\ref{eq:bse},
which is denoted as BSE-TDA.
TDA has been used in TDHF and TDDFT to overcome the triplet instability problem\cite{peachOvercomingLowOrbital2012,peachTripletInstabilityTDDFT2013}.
Recent works\cite{rangelAssessmentLowlyingExcitation2017} has shown that BSE-TDA
also provides an improved accuracy over BSE. \\

Eq.\ref{eq:bse} is analogous to the Casida equation\cite{casidaTimeDependentDensityFunctional1995,ullrichTimeDependentDensityFunctionalTheory2011} in TDDFT.
The only difference is that the BSE kernel replaces the XC kernel.
Thus, Eq.\ref{eq:bse} can be solved by the canonical Davidson algorithm\cite{stratmannEfficientImplementationTimedependent1998,davidsonIterativeCalculationFew1975}
with a $\mathcal{O}(N^4)$ scaling. \\

In the BSE/$GW$ approach,
the input orbital energies in Eq.\ref{eq:a} are QP energies from a $GW$ calculation.
In our BSE/LOSC approach,
LOSC orbital energies are directly used as the input orbital energies for BSE. \\

As shown in Section.1 of the Supporting Information
the scaling of adding the LOSC correction is only $\mathcal{O}(N^4)$,
which is the same as the scaling of solving the BSE equation in Eq.\ref{eq:bse}.
Therefore,
the overall scaling of the BSE/LOSC approach is only $\mathcal{O}(N^4)$ and is
computationally much more favorable than BSE/$G_0W_0$.
This opens the new possibility of the BSE formalism for describing excited
states of larger systems. \\

\section{\label{sec:computation}COMPUTATIONAL DETAILS}
We implemented the BSE/LOSC approach in the QM4D quantum chemistry package\cite{qm4d}
and applied it to calculate excitation energies of different systems.
For the Truhlar-Gagliardi test set\cite{hoyerMulticonfigurationPairDensityFunctional2016},
the aug-cc-pVTZ basis set\cite{dunningGaussianBasisSets1989, kendallElectronAffinitiesFirst1992} was used,
except that the aug-cc-pVDZ basis set\cite{dunningGaussianBasisSets1989, kendallElectronAffinitiesFirst1992} was used for naphthalene, pNA and DMABN.
B-TCNE was excluded because of the computational cost.
For Stein CT test set\cite{steinReliablePredictionCharge2009},
the cc-pVDZ\cite{dunningGaussianBasisSets1989} basis set was used.
The experiment values in the gas phase\cite{steinReliablePredictionCharge2009}
were taken as references.
For the test of Rydberg excitation energies of B$^+$, Be, Mg and Al$^+$,
the aug-cc-pVQZ basis set\cite{dunningGaussianBasisSets1989, kendallElectronAffinitiesFirst1992} was used.
Reference values were taken from Ref.\citenum{xuTestingNoncollinearSpinFlip2014}.
BSE/LOSC and BSE/$G_0W_0$ calculations were performed with QM4D,
TDDFT calculations were performed with GAUSSIAN16 A.03 software\cite{g16}.
QM4D uses Cartesian basis sets and the resolution of identity (RI) technique\cite{weigendAccurateCoulombfittingBasis2006,renResolutionofidentityApproachHartree2012,eichkornAuxiliaryBasisSets1995}
to compute two-electron integrals.
All basis sets and corresponding fitting basis sets were taken from the Basis Set Exchange\cite{fellerRoleDatabasesSupport1996,schuchardtBasisSetExchange2007,pritchardNewBasisSet2019}. \\

\section{\label{sec:results}RESULTS}

\subsection{\label{subsec:tg}Truhlar-Gagliardi test set}
\FloatBarrier

\fontsize{9}{11}\selectfont{
\begin{longtable}{@{\extracolsep{\fill}}cccccccc}
    \caption{\textnormal{Mean absolute errors (MAEs) and mean signed errors (MSEs) of excitation energies in Truhlar-Gagliardi test set obtained from BSE/LOSC,
    BSE/$G_0W_0$, TDDFT, BSE/ev$GW$, BSE-TDA/LOSC, BSE-TDA/$G_0W_0$, TDDFT-TDA and BSE-TDA/ev$GW$ based on HF, BLYP, PBE, B3LYP and PBE0.
    All values in eV.
    Geometries were taken from Ref.\citenum{hoyerMulticonfigurationPairDensityFunctional2016}.
    Reference values for pNA and DMABN were taken from Ref.\citenum{guiAccuracyAssessmentGW2018} and for remaining molecules were taken from Ref.\citenum{verilQUESTDBDatabaseHighly2021}.
    The aug-cc-pVDZ basis set was used for naphthalene, pNA and DMABN, and the aug-cc-pVTZ basis set was used for the remaining systems.
    B-TCNE was excluded considering the computational cost. Total MAEs and total MSEs were calculated by averaging all systems with equal weights.
    MAEs and MSEs for valence, Rydberg and CT excitations can be found in Table.S6 in the Supporting Information.}\label{tab:tgset}}\\\toprule
    & & \multicolumn{2}{c}{total} & \multicolumn{2}{c}{singlet} & \multicolumn{2}{c}{triplet}  \\
    \cmidrule(l{0.5em}r{0.5em}){3-4} \cmidrule(l{0.5em}r{0.5em}){5-6} \cmidrule(l{0.5em}r{0.5em}){7-8}
    & & MAE & MSE & MAE & MSE & MAE & MSE
\\\hline
\endfirsthead
\caption{Continued}\\\toprule
& & \multicolumn{2}{c}{total} & \multicolumn{2}{c}{singlet} & \multicolumn{2}{c}{triplet}  \\
\cmidrule(l{0.5em}r{0.5em}){3-4} \cmidrule(l{0.5em}r{0.5em}){5-6} \cmidrule(l{0.5em}r{0.5em}){7-8}
& & MAE & MSE & MAE & MSE & MAE & MSE
\\\hline
\endhead
\bottomrule
\endlastfoot
        BLYP  & BSE/LOSC         & 1.02  & -1.02 & 0.90    & -0.90 & 1.42    & -1.38 \\
              & BSE-TDA/LOSC     & 0.82  & -0.80 & 0.75    & -0.74 & 1.03    & -0.98 \\
        PBE   & BSE/LOSC         & 1.04  & -1.03 & 0.94    & -0.93 & 1.35    & -1.29 \\
              & BSE-TDA/LOSC     & 0.83  & -0.80 & 0.79    & -0.76 & 0.96    & -0.90 \\
        B3LYP & BSE/LOSC         & 0.62  & -0.56 & 0.51    & -0.44 & 0.96    & -0.89 \\
              & BSE-TDA/LOSC     & 0.46  & -0.38 & 0.41    & -0.30 & 0.63    & -0.56 \\
        PBE0  & BSE/LOSC         & 0.54  & -0.45 & 0.46    & -0.34 & 0.82    & -0.73 \\
              & BSE-TDA/LOSC     & 0.39  & -0.27 & 0.35    & -0.20 & 0.50    & -0.41 \\
        HF    & BSE/$G_0W_0$     & 0.89  & 0.87  & 0.88    & 0.85  & 0.93    & 1.05  \\
              & BSE-TDA/$G_0W_0$ & 0.95  & 0.93  & 0.92    & 0.89  & 1.06    & 1.17  \\
        BLYP  & BSE/$G_0W_0$     & 1.53  & -1.48 & 1.42    & -1.36 & 1.88    & -1.92 \\
              & BSE-TDA/$G_0W_0$ & 1.34  & -1.19 & 1.32    & -1.26 & 1.42    & -0.83 \\
        PBE   & BSE/$G_0W_0$     & 1.48  & -1.32 & 1.39    & -1.34 & 1.76    & -1.17 \\
              & BSE-TDA/$G_0W_0$ & 1.32  & -1.16 & 1.29    & -1.23 & 1.41    & -0.81 \\
        B3LYP & BSE/$G_0W_0$     & 1.11  & -0.90 & 1.01    & -0.92 & 1.43    & -0.68 \\
              & BSE-TDA/$G_0W_0$ & 0.97  & -0.76 & 0.92    & -0.83 & 1.16    & -0.43 \\
        PBE0  & BSE/$G_0W_0$     & 1.00  & -0.78 & 0.90    & -0.80 & 1.35    & -0.57 \\
              & BSE-TDA/$G_0W_0$ & 0.88  & -0.66 & 0.82    & -0.71 & 1.09    & -0.33 \\
        HF    & TDDFT            & 1.55  & -0.56 & 0.69    & 0.51  & 4.48    & -4.14 \\
              & TDDFT-TDA        & 0.78  & 0.46  & 0.82    & 0.69  & 0.63    & -0.21 \\
        BLYP  & TDDFT            & 0.62  & -0.59 & 0.68    & -0.65 & 0.40    & -0.46 \\
              & TDDFT-TDA        & 0.57  & -0.52 & 0.64    & -0.59 & 0.32    & -0.34 \\
        PBE   & TDDFT            & 0.59  & -0.56 & 0.65    & -0.61 & 0.40    & -0.46 \\
              & TDDFT-TDA        & 0.54  & -0.48 & 0.60    & -0.54 & 0.30    & -0.32 \\
        B3LYP & TDDFT            & 0.40  & -0.33 & 0.40    & -0.31 & 0.41    & -0.43 \\
              & TDDFT-TDA        & 0.32  & -0.23 & 0.36    & -0.25 & 0.18    & -0.19 \\
        PBE0  & TDDFT            & 0.36  & -0.27 & 0.32    & -0.21 & 0.51    & -0.52 \\
              & TDDFT-TDA        & 0.25  & -0.15 & 0.28    & -0.14 & 0.16    & -0.17 \\
        HF    & BSE/ev$GW$       & 0.65  & 0.62  & 0.71    & 0.68  & 0.43    & 0.43  \\
              & BSE-TDA/ev$GW$   & 0.87  & 0.85  & 0.83    & 0.80  & 1.00    & 1.10  \\
        BLYP  & BSE/ev$GW$       & 0.58  & -0.55 & 0.46    & -0.42 & 0.98    & -0.97 \\
              & BSE-TDA/ev$GW$   & 0.53  & -0.24 & 0.43    & -0.26 & 0.88    & -0.04 \\
        PBE   & BSE/ev$GW$       & 0.57  & -0.54 & 0.45    & -0.42 & 0.98    & -0.97 \\
              & BSE-TDA/ev$GW$   & 0.52  & -0.24 & 0.42    & -0.26 & 0.87    & -0.05 \\
        B3LYP & BSE/ev$GW$       & 0.54  & -0.52 & 0.44    & -0.40 & 0.90    & -0.88 \\
              & BSE-TDA/ev$GW$   & 0.51  & -0.21 & 0.42    & -0.24 & 0.83    & 0.03  \\
        PBE0  & BSE/ev$GW$       & 0.52  & -0.49 & 0.42    & -0.37 & 0.87    & -0.84 \\
              & BSE-TDA/ev$GW$   & 0.49  & -0.19 & 0.39    & -0.22 & 0.81    & 0.06
\end{longtable}
}

\FloatBarrier

We first examine the performance of the BSE/LOSC approach for predicting
excitation energies of systems in Truhlar-Gagliardi test set.
This test set contains 18 valence excitations as well as two Rydberg
excitations and two CT excitations.
The mean absolute errors (MAEs) and mean signed errors (MSEs) of excitation energies obtained from BSE/LOSC,
BSE/$G_0W_0$, TDDFT, BSE/ev$GW$, BSE-TDA/LOSC, BSE-TDA/$G_0W_0$, TDDFT-TDA and BSE-TDA/ev$GW$ with HF,
BLYP, PBE, B3LYP and PBE0 are listed in Table.\ref{tab:tgset}.
It shows that BSE/$G_0W_0$ has relatively large errors.
The MAEs of BSE/$G_0W_0$ with hybrid functionals are around $0.9$ \,{eV}
and of BSE/$G_0W_0$ with GGA functionals can even exceed $1.3$ \,{eV}.
The BSE/LOSC approach significantly outperforms BSE/$G_0W_0$ with both GGA and
hybrid functionals.
The MAEs of BSE/LOSC are reduced by around $0.4$ \,{eV} compared with BSE/$G_0W_0$.
TDDFT@PBE0 provides a small MAE of $0.28$ \,{eV},
which agrees with results from Ref.\citenum{hoyerMulticonfigurationPairDensityFunctional2016}.
BSE/ev$GW$ provides the largely reduced starting point dependence and only slightly larger MAEs than TDDFT with hybrid functionals.
We also find that using the TDA greatly improves the accuracy of BSE/LOSC in the calculation for this test set.
MAEs of BSE-TDA/LOSC are reduced by $0.1$ \,{eV} to $0.2$ \,{eV} compared with BSE/LOSC.
As can be seen in Table.\ref{tab:tgset},
using TDA leads to increased excitation energies and improves the accuracy for triplet excitations in BSE/LOSC and BSE/$G_0W_0$,
which largely underestimate triplet excitation energies.
However, one should be careful when using TDA in the BSE/$GW$ approach.
As shown in the present work and Ref.\citenum{jacqueminBetheSalpeterFormalism2017},
using TDA in BSE/ev$GW$ leads to similar or worse results and provides minor effects for singlet calculations.
In addition,
as shown in recent studies using TDA in BSE/$GW$ can lead to blue-shifts in nanosized systems\cite{faberManybodyGreenFunction2013,roccaInitioCalculationsOptical2010,ducheminShortRangeLongRangeChargeTransfer2012} and worse estimations for singlet-triplet energy gaps in organic molecules\cite{jacqueminBenchmarkBetheSalpeterTriplet2017}.
As shown in details in Section.2 in the Supporting Information,
BSE/LOSC and BSE/$G_0W_0$ largely underestimate valence excitation energies.
Excitation energies obtained from BSE-TDA/LOSC and BSE-TDA/$G_0W_0$ are always
higher than BSE/LOSC and BSE/$G_0W_0$ by $0.1$ \,{eV} to $0.7$ \,{eV},
which lead to smaller errors.
In BSE/LOSC,
BSE-TDA/LOSC@PBE0 provides the smallest MAE of $0.36$ \,{eV},
which is close to the accuracy of TDDFT-TDA with hybrid functionals and BSE/ev$GW$.
Besides the improved accuracy,
BSE/LOSC and BSE-TDA/LOSC approaches also reduce the starting point dependence
compared with BSE/$G_0W_0$ and BSE-TDA/$G_0W_0$.
However, there is still a large difference between using GGA and hybrid
functionals for predicting excitation energies in this test set. \\

\FloatBarrier

\subsection{\label{subsec:stein}Stein CT test set}
\FloatBarrier

\begin{table}
    \caption{\label{tab:steinset}Mean absolute errors (MAEs) and mean signed errors (MSEs) of CT excitation energies in Stein CT test set obtained
    from BSE/LOSC, BSE/$G_0W_0$, TDDFT, BSE-TDA/LOSC, BSE-TDA/$G_0W_0$ and
    TDDFT-TDA with HF, BLYP, PBE, B3LYP and PBE0, all values in eV\tnote{2,3}.
    References and geometries were taken from Ref.\citenum{steinReliablePredictionCharge2009}.
    Gas phase references were used.
    The cc-pVDZ basis set was used for all systems. }
    \begin{ruledtabular}
        \begin{tabular}{c|cccccccccc}
            & \multicolumn{2}{c}{HF} & \multicolumn{2}{c}{BLYP} & \multicolumn{2}{c}{PBE} & \multicolumn{2}{c}{B3LYP} & \multicolumn{2}{c}{PBE0} \\
            \cmidrule(l{0.5em}r{0.5em}){2-3}\cmidrule(l{0.5em}r{0.5em}){4-5} \cmidrule(l{0.5em}r{0.5em}){6-7} \cmidrule(l{0.5em}r{0.5em}){8-9} \cmidrule(l{0.5em}r{0.5em}){10-11}
            & MAE & MSE & MAE & MSE & MAE & MSE & MAE & MSE & MAE & MSE \\ \hline
            BSE/LOSC          &   &   & 0.50 & -0.36 & 0.50 & -0.38 & 0.45 & -0.39 & 0.42 & -0.37 \\
            BSE/$G_0W_0$      & 0.10 & -0.06 & 1.28 & -1.28 & 1.31 & -1.31 & 0.74 & -0.74 & 0.65 & -0.65 \\
            TDDFT             & 0.78 & 0.78 & 1.44 & -1.44 & 1.45 & -1.45 & 1.16 & -1.16 & 1.08 & -1.08 \\
            BSE-TDA/LOSC      &   &   & 0.56 & -0.27 & 0.57 & -0.29 & 0.46 & -0.32 & 0.43 & -0.31 \\
            BSE-TDA/$G_0W_0$  & 0.11 & -0.04 & 1.13 & -1.13 & 1.16 & -1.16 & 0.66 & -0.66 & 0.59 & -0.59 \\
            TDDFT-TDA         & 0.80  & 0.80 & 1.34 & -1.30 & 1.35 & -1.32 & 1.10 & -1.07 & 1.03 & -1.00\\
        \end{tabular}
    \end{ruledtabular}
\end{table}


We further investigate the performance of our BSE/LOSC approach on predicting
CT excitation energies by testing 12 CT systems in Stein' set.
The MAEs of predicting CT excitation energies obtained from BSE/LOSC,
BSE/$G_0W_0$, TDDFT, BSE-TDA/LOSC, BSE-TDA/$G_0W_0$ and TDDFT-TDA with HF, BLYP,
PBE, B3LYP and PBE0 are listed in Table.\ref{tab:steinset}.
It can be seen that TDDFT with both GGA and hybrid functionals fails to predict
CT excitation energies due to the wrong description for the long-range behavior.
The MAEs are around $1.4$ \,{eV} for TDDFT with GGA functionals and $1.1$ \,{eV}
for TDDFT with hybrid functionals.
Because of the correct long-range behavior,
BSE/$G_0W_0$ shows smaller MAEs compared with TDDFT.
BSE/$G_0W_0$ has MAEs around $1.3$ \,{eV} with GGA functionals and only around
$0.7$ \,{eV} with hybrid functionals.
BSE/$G_0W_0$@HF gives a very small MAE of $0.10$ \,{eV}.
Our BSE/LOSC approach provides further improvements over BSE/$G_0W_0$.
BSE/LOSC with both GGA functionals and hybrid functionals has a small MAE of $0.5$ \,{eV}.
However, the MAEs obtained from BSE/LOSC are larger than the MAE of $0.10$ \,{eV} obtained from BSE/ev$GW$ reported in Ref.\citenum{blaseChargetransferExcitationsMolecular2011}.
In addition to the improved accuracy over BSE/$G_0W_0$,
the starting point dependence is largely eliminated in the BSE/LOSC approach.
We also find that BSE-TDA/LOSC gives very close results to BSE/LOSC for predicting CT
excitation energies.
This observation is different from the results of Truhlar-Gagliardi test set,
where BSE-TDA/LOSC results are always better. \\

\FloatBarrier

\subsection{\label{subsec:rydberg}Rydberg excitations}
\FloatBarrier

\begin{table}
    \caption{\label{tab:rydberg}
    Mean absolute errors of Rydberg excitation energies of B$^+$, Be and Mg,
    all values in eV.
    References were taken from Ref.\citenum{xuTestingNoncollinearSpinFlip2014}.
    The aug-cc-pVQZ basis set was used.}
    \begin{ruledtabular}
        \begin{tabular}{c|ccccc}
                          & HF   & BLYP & PBE  & B3LYP & PBE0 \\
        \hline
        BSE/LOSC          &      & 0.88 & 0.69 & 0.74  & 0.56 \\
        BSE/$G_0W_0$      & 0.16 & 1.00 & 0.97 & 0.73  & 0.64 \\
        BSE/ev$GW$      & 0.15 & 0.65 & 0.65 & 0.59  & 0.44 \\
        TDDFT             & 0.92 & 1.17 & 1.03 & 0.89  & 0.80 \\
        BSE-TDA/LOSC      &      & 0.84 & 0.65 & 0.71  & 0.54 \\
        BSE-TDA/$G_0W_0$  & 0.15 & 1.00 & 0.97 & 0.70  & 0.61 \\
        TDDFT-TDA         & 0.90 & 1.16 & 1.02 & 0.88  & 0.79 \\
        BSE-TDA/ev$GW$      & 0.14 & 0.61 & 0.61 & 0.55  & 0.51 \\
        \end{tabular}
    \end{ruledtabular}
\end{table}

We further study the performance of our BSE/LOSC approach on predicting
Rydberg excitation energies by testing B$^+$, Be and Mg.
The MAEs of predicting Rydberg excitation energies obtained from BSE/LOSC,
BSE/$G_0W_0$, TDDFT, BSE/ev$GW$, BSE-TDA/LOSC, BSE-TDA/$G_0W_0$, TDDFT-TDA and BSE-TDA/ev$GW$ with HF, BLYP,
PBE, B3LYP and PBE0 are listed in Table.\ref{tab:rydberg}.
Similar to the CT excitation energies,
TDDFT also fails to predict Rydberg excitation energies.
The MAEs of TDDFT with GGA functionals or hybrid functionals are relatively large.
BSE/$G_0W_0$ provides improvements over TDDFT for both GGA and hybrid functionals,
where MAEs are reduced by $0.1$ \,{eV} to $0.2$ \,{eV}.
BSE/ev$GW$ provides accurate Rydberg excitation energies with the minimal starting point dependence.
Our BSE/LOSC approach provides the best accuracy for Rydberg excitation energies.
The MAEs of BSE/LOSC are further reduced by $0.1$ \,{eV} to $0.2$ \,{eV} compared with
BSE/$G_0W_0$.
The accuracy of BSE/LOSC with hybrid functionals is comparable to the BSE/ev$GW$ level.
BSE-TDA/LOSC yields very close results to BSE/LOSC for Rydberg excitations,
which is similar to the case of CT excitations.

\FloatBarrier

\section{\label{sec:conclusions}CONCLUSIONS}
In summary,
we applied LOSC in BSE to calculate excitation energies of molecular systems.
In the BSE/LOSC approach,
the LOSC correction is added in the post-SCF manner to correct the KS orbital energies.
Then the LOSC orbital energies are directly used in BSE.
The BSE-TDA/LOSC can be obtained by neglecting the $\mathbf{B}$ matrix in the BSE calculation.
BSE/LOSC was first examined on predicting excitation energies
in Truhlar-Gagliardi test set.
We showed that BSE/LOSC significantly outperforms BSE/$G_0W_0$ and BSE-TDA/LOSC
provides further improvements.
BSE-TDA/LOSC with hybrid functionals provides the comparable accuracy to TDDFT
for predicting excitation energies in this set.
Then we showed that BSE/LOSC predicts accurate CT excitation energies in Stein CT test set.
BSE/LOSC provides considerable improvements over BSE/$G_0W_0$ and largely
eliminates the starting point dependence.
We also showed that the BSE/LOSC approach describes Rydberg
excitations well by testing atomic Rydberg excitation energies.
Therefore, the BSE/LOSC approach greatly outperforms BSE/$G_0W_0$ for predicting
valence, CT and Rydberg excitation energies.
Besides the improved accuracy,
our BSE/LOSC approach only scales as $\mathcal{O}(N^4)$,
which is much more computationally favorable than BSE/$G_0W_0$.
The BSE/LOSC approach is expected to extend the applicability of the BSE
formalism for large system.

\section*{SUPPORTING INFORMATION}
See the Supporting Information for the scaling analysis of LOSC, and the numerical results of excitation energies for Truhlar-Gagliardi test set, Stein charge transfer test set and Rydberg excitations of atoms.

\begin{acknowledgements}
    ACKNOWLEDGMENTS: J. L. and Y. J. acknowledge the support from the National
    Institute of General Medical Sciences of the National Institutes of
    Health under award number R01-GM061870. N.Q.S and W.Y. acknowledge the support
    from the National Science Foundation (grant no. CHE-1900338).
\end{acknowledgements}

\section*{Data Availability Statement}
The data that support the findings of this study are available from the corresponding author
upon reasonable request.

\bibliography{ref,software}

\end{document}